\documentclass[twocolumn,english,aps,prl,floats]{revtex4-2}
\usepackage[T1]{fontenc}
\usepackage[latin9]{inputenc}
\usepackage{babel}
\usepackage{bm}
\usepackage{amsmath}
\usepackage{amssymb}
\usepackage{graphicx}
\usepackage{esint}
\usepackage[bookmarks=false,
 breaklinks=false,pdfborder={0 0 1},backref=false,colorlinks=false]
 {hyperref}

\makeatletter
\usepackage{babel}

\usepackage{siunitx}
\DeclareSIUnit\angstrom{\text{\AA}}
\usepackage{xcolor}

\@ifundefined{textcolor}{}{%
 \definecolor{BLACK}{gray}{0}
 \definecolor{WHITE}{gray}{1}
 \definecolor{RED}{rgb}{1,0,0}
 \definecolor{GREEN}{rgb}{0,1,0}
 \definecolor{BLUE}{rgb}{0,0,1}
 \definecolor{CYAN}{cmyk}{1,0,0,0}
 \definecolor{MAGENTA}{cmyk}{0,1,0,0}
 \definecolor{YELLOW}{cmyk}{0,0,1,0}
}

\usepackage{ifpdf}\usepackage{bm}

\makeatother

\begin{document}
\title{Skyrmion Cyclotron Resonance in Ferrimagnets}
\author{Eugene M. Chudnovsky and Dmitry A. Garanin}
\affiliation{Physics Department, Herbert H. Lehman College and Graduate School,
The City University of New York, 250 Bedford Park Boulevard West,
Bronx, New York 10468-1589, USA }
\date{\today}
\begin{abstract}
We show that a resonance due to gyroscopic motion of skyrmions, conceptually
similar to the electron cyclotron resonance in metals, can be excited
in a ferrimagnetic film by a spin current or microwaves. It must permit
unambiguous measurement of the skyrmion mass for which a universal
expression depending solely on the exchange interaction between spins
belonging to two different ferrimagnetic sublattices is derived. The
dependence of the skyrmion cyclotron frequency on parameters is computed
for a TM/RE ferrimagnet, using CoGd as an example. The cyclotron frequency
exhibits a dip near the angular momentum compensation point, where
it hybridizes with the ferromagnetic resonance. The skyrmion cyclotron
mode is studied for individual skyrmions and for skyrmion lattices,
where the effect must be strong enough to be observed in microwave
and spin-current experiments. 
\end{abstract}
\maketitle
In this Letter, we address two interconnected problems that are at
the forefront of theoretical and experimental research on magnetism
of solids: spin excitations in ferrimagnets near the angular momentum
compensation point, and the mass of a ferrimagnetic skyrmion. Recent
interest to ferrimagnets has been generated by observation that while
possessing a nonzero magnetization they can exhibit fast dynamics
similar to that of compensated antiferromagnets \citep*{Binder-PRB2006,Arena-PRApplied2017,Siddiqui-PRB2018,Ivanov-review,Bonfiglio-PRB2019,Kim-Nature2020,DavydovaJOP2020,Yurlov-PRB2021,Joo-materials2021,Chanda-PRB2021,Kim-NatMat2022,Guo-PRB2022,Zhang-PRB2022,Chen-Materials2025,Moreno-PRB2025,Ciccarelli-AP2025}.
It stems from the fact that non-equal gyromagnetic factors $g_{i}$
of antiferromagnetically coupled, ferromagnetically ordered sublattices
permit a nonzero average magnetic moment, $\sum_{i}g_{i}\mu_{B}\langle{\bf S}_{i}\rangle$
when the average angular momentum $\sum_{i}\hbar\langle{\bf S}_{i}\rangle$
associated with the atomic spins of the sublattices $\mathbf{S}_{i}$
turns zero at a certain temperature or composition of the ferrimagnet.
Hardening of the magnetic resonance, and spin waves in general, on
approaching the angular momentum compensation point in ferrimagnets
has been demonstrated by several theoretical studies, see, e.g., \onlinecite{Lin-PRB1988,Zhang-JPhys1997,Karchev-JPhys2008,Okuno-APLExpress2019,Haltz-PRB2022,Sanchez-PRB2025,DG-EC-PRB2026},
and confirmed by numerous experiments, see, e.g., \onlinecite{Pardavi-JMMM2000,Binder-PRB2006,Stanciu-PRB2006,Okuno-APLExpress2019,Kim-Nature2020}.

Likewise, the interest in skyrmions, besides their mathematical beauty,
has been inspired by their potential for topologically protected information
technology \citep{APL-2021}. Various inhomogeneous magnetic phases,
including magnetic domains and skyrmion lattices (SkL), have been
observed in ferrimagnetic films and racetracks \citep{Caretta-NatNano2018,Woo-NatCom2018,Hirata-NatNano2019,Kim-PRL2019}.
Excitation modes of individual skyrmions and SkL have been studied
in ferromagnets (see, e.g., Refs.\ \onlinecite{DA-RJ-EC-PRB2020,DG-EC-PRB2025},
and references therein), but not so much in ferrimagnets. The dynamics
of skyrmions largely depends on whether it is massive or massless.
The question of the skyrmion mass has been one of the most controversial
in skyrmion physics \citep{Makhfudz-PRL2012,Buttner-Nature2015,Shiino2017,Lin-PRB2017,Psaroudaki-PRX2017,Kravchuk-PRB2018,Li-PRB2018}.
It has been rigorously demonstrated that in a 2D ferromagnetic film
the inertial mass associated with the motion of the center of the
topological charge of the skyrmion is zero \citep{Tchernyshyov,EC-DA-EPL2025}.
It was suggested that a nonzero skyrmion mass can arise from its confinement
in a nanoring \citep{Liu-MMM2020} or nanotrack \citep{EC-DA-EPL2025},
or it can be generated by the interaction with other degrees of freedom,
such as phonons \citep{Capic-PRB2020}.

In this Letter we show that a finite skyrmion mass is generated in
a ferrimagnet by the inter-sublattice exchange interaction alone.
The mass is independent of the atomic spins and spin densities of
the sublattices. The finite mass of the skyrmion generates its gyroscopic
motion at a frequency $\Omega=(J'/\hbar)(S-c\Sigma)$, where $J'>0$
is the strength of the antiferromagnetic exchange interaction between
spins $S$ and $\Sigma$ of the sublattices in ${\cal H}_{\mathrm{int}}=J'\sum_{i}{\bf S}_{i}\cdot\bm{\Sigma}_{i}$,
and $c$ is the concentration of spins $\Sigma$. We call $\Omega$
the frequency of the skyrmion cyclotron resonance (SCR), and confirm
its analytical value by numerical computation of spin-current and
microwave power absorption that uses a microscopic lattice-spin model.
The SCR frequency decreases on approaching the angular momentum compensation
point $c\rightarrow c^{*}=S/\Sigma$, where it hybridizes with the
FMR frequency, pushing it down. (Having a TM/RE ferrimagnet in mind
we assume $S<\Sigma$). Our calculations have been performed for individual
skyrmions and SkL using parameters of a CoGd ferrimagnet, for which
our predictions can be tested in experiments. Measurements of the
SCR must allow unambiguous determination of the evasive skyrmion mass,
similar to how the effective electron mass in metals was obtained
from measurements of the Azbel-Kaner electron cyclotron resonance
\citep{ECR}.

\begin{figure}
\centering{}\includegraphics[width=1\linewidth]{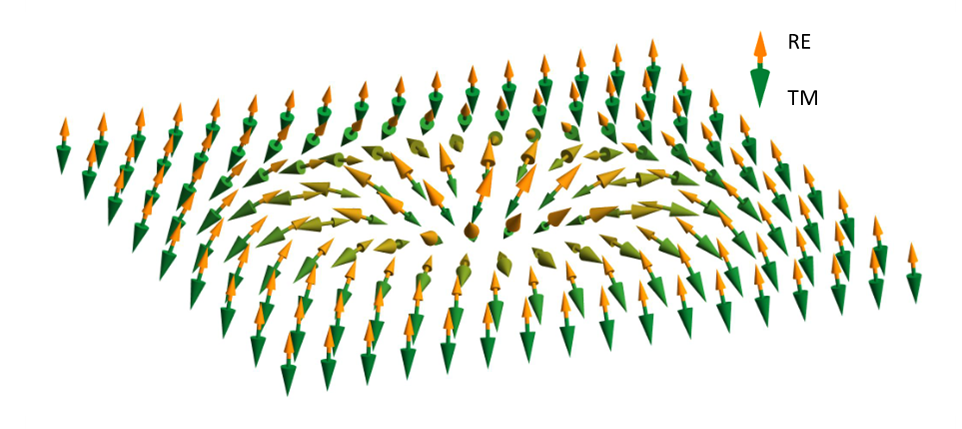}
\caption{Two-sublattice skyrmion in a TM/RE ferrimagnet. Far from the skyrmion's
core, TM spins are directed down (green) and RE spins are directed
up (orange).}
\label{skyrmion} 
\end{figure}

A ferrimagnetic skyrmion is shown in Fig.\ \ref{skyrmion}. A classical
2D field-model counterpart of ${\cal H}_{\mathrm{int}}=J'\sum_{i}{\bf S}_{i}\cdot\bm{\Sigma}_{i}$
on a square lattice of spacing $a$, suitable for its study, is $U_{\mathrm{int}}=J'a^{2}\int d^{2}r\bm{\rho}_{S}({\bf r})\cdot\bm{\rho}_{\Sigma}({\bf r})$,
where $\bm{\rho}_{S}$ and $\bm{\rho}_{\Sigma}$ are spin densities
of the sublattices $\rho_{S}=S/a^{2}$ and $\rho_{\Sigma}=c\Sigma/a^{2}$
respectively. We assume \citep{Panigrahy2022,Nowak-PRB2023,Lau2025}
that they are given by rigid skyrmion solutions $\bm{\rho}_{S}=\rho_{S}{\bf f}({\bf r}-{\bf R}_{S})$
and $\bm{\rho}_{\Sigma}=\rho_{\Sigma}{\bf f}({\bf r}-{\bf R}_{\Sigma})$
centered at ${\bf R}_{S}$ and ${\bf R}_{\Sigma}$ respectively, separated
by the distance ${\bf d}={\bf R}_{S}-{\bf R}_{\Sigma}$, which is
very small compared to the size of the skyrmion $\lambda$. This yields
\begin{equation}
U_{\mathrm{int}}=J'a^{2}\rho_{S}\rho_{\Sigma}\int d^{2}r{\bf f}({\bf r})\cdot{\bf f}({\bf r}+{\bf d})\cong\frac{1}{2}J'a^{2}\mathcal{D}\rho_{S}\rho_{\Sigma}{\bf d}^{2},
\end{equation}
where $\mathcal{D}=\frac{1}{2}\int dxdy\left[\partial{\bf f}({\bf r})/\partial{\bf r}\right]^{2}$
equals $4\pi|Q|$ for a Belavin-Polyakov pure-exchange skyrmion \citep{BP}
with a topological charge $Q$. Notice that the exchange energy associated
with the separation of the centers of sublattice skyrmions scales
as $J'(d/a)^{2}$, so that in practice one should always expect $d\ll a$.

In the absence of external forces, the dynamics of the two skyrmions
moving at velocities ${\bf V}_{S}$ and ${\bf V}_{\Sigma}$ satisfies
Thiele equations \citep{Thiele} augmented by the spin-current terms
\begin{eqnarray}
-{\bf G}_{S}\times\left({\bf V}_{S}-\mathbf{v}_{s,S}\right) & = & \mathbf{F}_{S}+{\bf F}_{S\Sigma},\nonumber \\
{\bf G}_{\Sigma}\times\left({\bf V}_{\Sigma}-\mathbf{v}_{s,\Sigma}\right) & = & \mathbf{F}_{\Sigma}+{\bf F}_{\Sigma S},\label{Thiele12}
\end{eqnarray}
where ${\bf G}_{S,\Sigma}=4\pi\hbar\left|Q\right|\rho_{S,\Sigma}\mathbf{e}_{z}$
are the so-called gyrovectors, $\mathbf{v}_{s,S}$ and $\mathbf{v}_{s,\Sigma}$
are spin currents in the TM and RE sublattices, $\mathbf{F}_{S}$
and $\mathbf{F}_{\Sigma}$ are external forces acting on the $S$-
and $\Sigma$-skyrmions, while the interaction force is given by ${\bf F}_{S\Sigma}=-{\bf F}_{\Sigma S}=-\partial U_{\mathrm{int}}/\partial{\bf R}_{S}=\partial U_{\mathrm{int}}/\partial{\bf R}_{\Sigma}=-J'a^{2}\mathcal{D}\rho_{S}\rho_{\Sigma}{\bf d}$.
In the simplest case without external forces and spin currents, for
the velocity ${\bf V}=\dot{{\bf R}}$ of the coordinate of the center
of mass of the two skyrmions, ${\bf R}=\left({\bf R}_{S}+{\bf R}_{\Sigma}\right)/2$,
and the acceleration $\dot{{\bf V}}$ from Eq.\ (\ref{Thiele12})
resolved for the velocities one obtains 
\begin{eqnarray}
{\bf V} & = & -\frac{1}{2}J'a^{2}\mathcal{D}\rho_{S}\rho_{\Sigma}\left(\frac{1}{G_{S}}+\frac{1}{G_{\Sigma}}\right)\mathbf{e}_{z}\times{\bf d},\label{V-d}\\
\dot{{\bf V}} & = & -J'a^{2}\mathcal{D}\rho_{S}\rho_{\Sigma}\left(\frac{1}{G_{S}}-\frac{1}{G_{\Sigma}}\right)\mathbf{e}_{z}\times{\bf V}.\label{dotV}
\end{eqnarray}
The second equation describing the motion of the free skyrmion is
obtained by the differentiation of the first one, expressing ${\bf \dot{d}}$
through the velocities, expressing the latter through ${\bf d}$ again
using Eq.\ (\ref{Thiele12}), and using the first equation to eliminate
${\bf d}$. This equation can be rewritten as $\mathbf{\dot{V}}=\boldsymbol{\Omega}\times{\bf V}$,
where the cyclotron frequency of the skyrmion is given by a very simple
expression
\begin{equation}
\boldsymbol{\Omega}=\frac{J'}{\hbar}\frac{\mathcal{D}}{4\pi\left|Q\right|}\left(S-c\Sigma\right)\mathbf{e}_{z}\Rightarrow\frac{J'}{\hbar}\left(S-c\Sigma\right)\mathbf{e}_{z},\label{Omega_cyclotron}
\end{equation}
if $\mathcal{D}=4\pi|Q|$. The corresponding cyclotron radius is $R_{c}=V/\left|\Omega\right|$
and diverges at the angular-momentum compensation point. From Eq.
(\ref{V-d}) one obtains $V=(1/2)(J'/\hbar)\mathcal{D}/\left(4\pi\left|Q\right|\right)\left(S+c\Sigma\right)d$,
thus $R_{c}=(d/2)(S+c\Sigma)/|S-c\Sigma|$. As $d$ is very small,
one needs to come close to the angular-momentum compensation point
to have a significant $R_{c}$.

To identify the skyrmion mass, one should include the external force
${\bf F}={\bf F}_{S}+{\bf F}_{\Sigma}$. With a slightly more effort,
for $\mathbf{F}=\mathrm{const}$ one obtains
\begin{equation}
M\dot{V}=({\bf G}_{S}-{\bf G}_{\Sigma})\times{\bf V}+{\bf F}=M\boldsymbol{\Omega}\times{\bf V}+{\bf F},\label{Equation_of_motion}
\end{equation}
where $M=(4\pi\hbar Q)^{2}/\left(J'a^{2}\mathcal{D}\right)$ is the
skyrmion mass. For $D=4\pi|Q|$ it equals 
\begin{equation}
M=\frac{4\pi\hbar^{2}|Q|}{J'a^{2}}.\label{skyrniom_mass}
\end{equation}
It must be of the order of the proton mass for a single 2D atomic
layer of CoGd ferrimagnet. Experimentally, the skyrmion mass can be
found as 
\begin{equation}
M=\frac{G_{S}-G_{\Sigma}}{\Omega}\label{M_via_cyclotron}
\end{equation}
from the cyclotron-resonance experiment driven by the spin current
or microwave radiation, where the difference of gyrovectors $G_{S}-G_{\Sigma}$
plays the role of the magnetic field acting on an electric charge.
Remarkably, $M$ doesn't depend on the spins (cf. Ref. \citep{Nowak-PRB2023}).
The proportionality of the skyrmion mass to $|Q|$ points to its topological
origin. One should keep in mind, however, that the applicability of
Eq.\ (\ref{skyrniom_mass}) depends on two assumptions. The first
is the condition of the rigidity of the sublattice skyrmions, which
requires the exchange interaction $J$ responsible for the skyrmion
texture within each ferromagnetically ordered sublattice to be large
compared to the intersublattice exchange interaction $J'$. The second
is that the skyrmion is formed by many spins belonging to each sublattice,
which requires $c\gg(a/\lambda)^{2}$, where $\lambda$ is the size
of the skyrmion. 

Equation\ (\ref{Equation_of_motion}) should be augmented by the
inhomogeneous terms due to spin current and time derivatives of the
external force and the spin current. The addition of weak dissipation
to the Thiele equations (\ref{Thiele12}) adds a small imaginary part,
$i\varLambda(J'/\hbar)(S+c\Sigma)$, to Eq.\ (\ref{Omega_cyclotron}),
where $\varLambda\equiv\alpha\mathcal{D}/\left(4\pi\left|Q\right|\right)$
and $\alpha\ll1$ is the phenomenological damping constant in the
Landau-Lifshitz equation for the magnetization. The mass gets modified
by a factor $1+\varLambda^{2}$. 

Notice that the lattice spacing $a$ that we used in the definitions
of spin densities and gyrovectors has canceled out from the formula
for $\Omega$. For a film consisting of $N$ atomic layers, both $M$
and the gyrovectors become proportional to $N$. Consequently, the
expression (\ref{Omega_cyclotron}) for the skyrmion cyclotron frequency,
derived for a single 2D layer, remains unchanged in a film of finite
thickness.

To test our analytical theory, we studied numerically excitations
in a $2D$ model of a TM/RE ferrimagnet with skyrmions, described
by the Hamiltonian 
\begin{eqnarray}
{\cal H} & = & -\frac{1}{2}\sum_{ij}J_{ij}{\bf S}_{i}\cdot{\bf S}_{j}+J'\sum_{i}{\bf S}_{i}\cdot p_{i}\bm{\Sigma}_{i}-\frac{D}{2}\sum_{i}S_{i,z}^{2}\nonumber \\
 & + & A\sum_{i}\left[({\bf S}_{i}\times{\bf S}_{i+\delta_{x}})_{x}+{\bf S}_{i}\times{\bf S}_{i+\delta_{y}})_{y}\right].\label{Ham}
\end{eqnarray}
Here $J_{ij}$ is the ferromagnetic nearest-neighbor exchange within
the square TM-sublattice of spins ${\bf S}_{i}$ with the coupling
constant $J>0$. The RE spins $\bm{\Sigma}_{i}$ occupy a similar
square lattice and are coupled to the TM spins with the coupling constant
$J'>0$. These spins are diluted, which is taken into account by the
occupation factors $p_{i}=0,1$, so that $\left\langle p_{i}\right\rangle =c$.
Other terms are the easy-axis anisotropy and the Dzialoshinskii-Moriya
interaction (DMI) within the TM sublattice. Having CoGd in mind, we
chose $S=3/2$, $\Sigma=7/2$, $J'/J=0.2$, $D/J=0.03$, and $A/J=0.1$.
To the Hamiltonian above, one can add the Zeeman term due to the external
magnetic field, as well as the time-dependent microwave (MW) field.
In this model, Gd spins are slaves of the Co spins. Consequently,
when the separation of the centers of Co and Gd skyrmions is small
compared to the lattice spacing, one can still expect the condition
of the rigidity of the sublattice skyrmions to be roughly satisfied.
The dynamics of the lattice spins is described by the Landau-Lifshitz
(LL) equation, augmented by the spin-current term:
\begin{equation}
\hbar\frac{\partial\mathbf{S}_{i}}{\partial t}+\hbar\left(\mathbf{v}_{s,S}\cdot\nabla\right)\mathbf{S}_{i}=\mathbf{S}_{i}\times\mathbf{H}_{\mathrm{eff},S,i},\label{LL}
\end{equation}
where $\mathbf{H}_{\mathrm{eff},S,i}=-\partial\mathcal{H}/\partial\mathbf{S}_{i}$
is the effective field (in the energy units) acting on the TM sublattice
and $\mathbf{v}_{s,S}$ is the spin current acting on the TM spins
$\mathbf{S}_{i}$. The action of this streaming term is translating
the whole set of $\mathbf{S}_{i}$ with the velocity $\mathbf{v}_{s,S}$.
The equation of motion for the RE spins is similar, only the spin-current
term is negligible because RE spins are due to the $f$-electrons
which are close to the atomic cores and effectively decoupled from
the charge carriers. We do not include the phenomenological damping
in the LL equation since the model shows a significant intrinsic damping
due to the RE disorder and/or temperature.

In the absence of skyrmions, the ferrimagnet described by the above
Hamiltonian possesses two excitation modes: the low-frequency acoustic
mode and the high-frequency optical mode. Various limits for these
modes have been worked out in Ref.\ \onlinecite{DG-EC-PRB2026}.
At $c\rightarrow0$, the acoustic mode tends to the FMR frequency
of the Co sublattice alone, $\hbar\omega_{\mathrm{LF}}/(SJ)=D/J=0.03$.
As $c$ increases, the frequency of this mode goes up. For $D\ll J'$,
at the angular-momentum compensation, $c=S/\Sigma$, one has $\hbar\omega_{\mathrm{HF,LF}}=S\left(\sqrt{DJ'}\pm D/2\right)$,
making $\hbar\omega_{\mathrm{LF}}/(SJ)\approx0.0625$ and $\hbar\omega_{\mathrm{HF}}/(SJ)\approx0.0925$.
At small $c$, the SCR frequency $\Omega$ of Eq.\ (\ref{Omega_cyclotron}),
divided by $SJ/\hbar$, is approximately $J'/J=0.2$. In real units
for CoGd ferrimagnet, $\Omega/\left(2\pi\right)\Rightarrow SJ'/(2\pi\hbar)$
must be in the ballpark of a few hundred $\mathrm{GHz}$ at small
$c$, going down on approach to compensation, $c\rightarrow S/\Sigma$,
where it may be accessible by experiments. Since the frequency of
the acoustic mode goes up on increasing $c$ from zero, while the
frequency of the cyclotron mode goes down, one should expect their
hybridization on approach to the angular momentum compensation. This
has been observed in the numerical experiment described below. Physically,
the mode coupling occurs because precession of spins in the acoustic
mode results in the circular motion of the skyrmion's top that is
the cyclotron motion. For this reason, the latter can be excited by
the MW radiation.

\begin{figure}
\begin{centering}
\includegraphics[width=8cm]{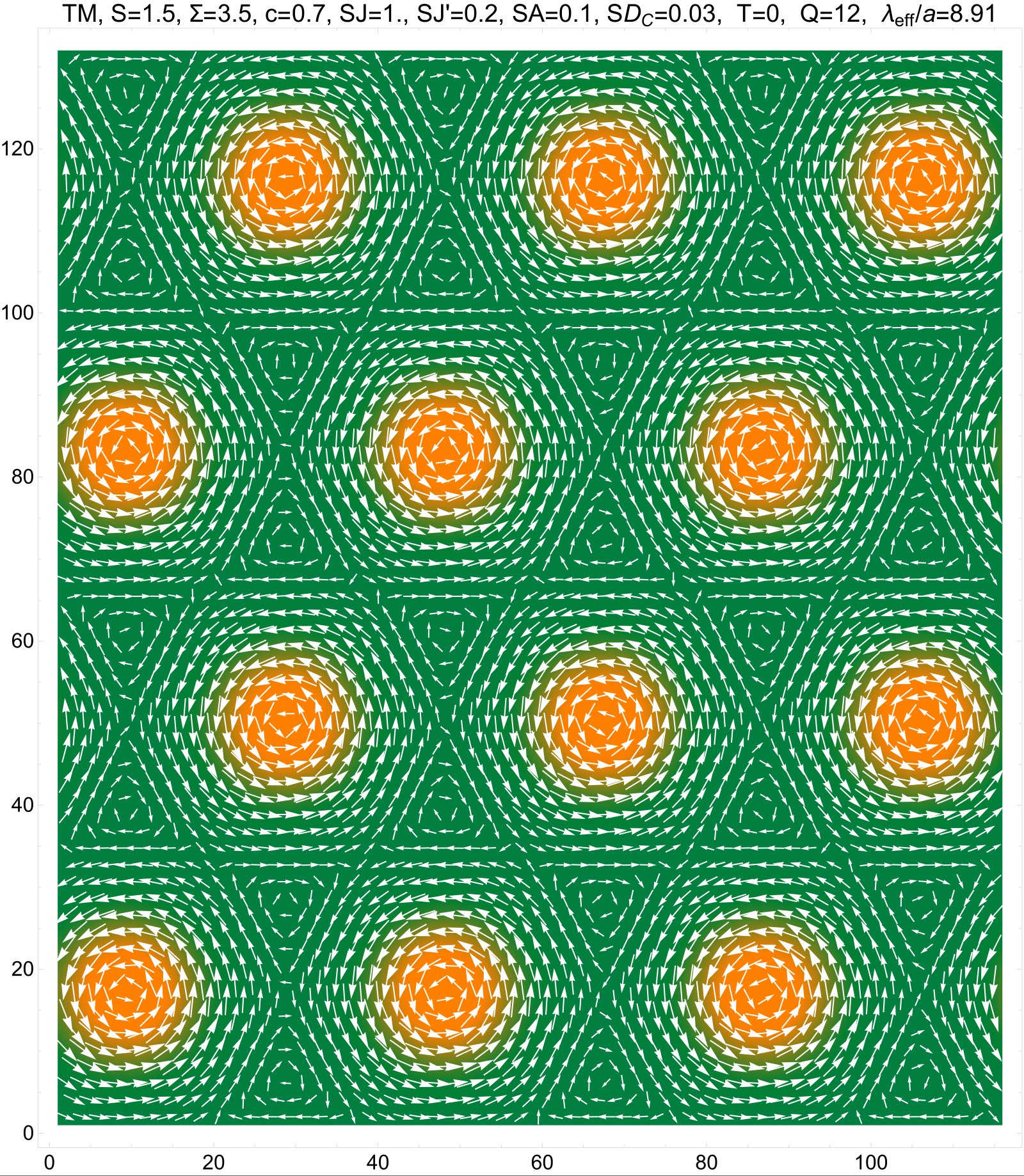}
\par\end{centering}
\caption{The system with a skyrmion lattice (TM spins are shown) for the chosen
model parameters and $c=0.7$. The skyrmion size $\lambda_{\mathrm{eff}}/a\approx8.9$
is defined according to Ref. \citep{caichugar2012}. }\label{Fig_SkL}

\end{figure}

To find the equilibrium states of the system with a single skyrmion
or a skyrmion lattice (SkL) at $T=0$, we performed the energy minimization
starting from the appropriate initial conditions (see, e.g., Ref.
\citep{gar25jpcm}). To find the modes' frequencies, we excited the
system with the sinc spin current or a sinc MW field directed along
the $x$ axis (perpendicular to the spins in the bulk and in the skyrmions'
cores), e.g., $\mathbf{v}_{s,S}(t)=\mathbf{e}_{x}v_{s}\sin\left(\omega_{\max}t\right)/\left(\omega_{\max}t\right)$.
The Fourier spectrum of this function is a constant up to the cutoff
frequency $\omega_{\max}$ and zero above it, so that it excites all
modes in the interval $0<\omega<\omega_{\max}$. The LL equation was
solved with the 5th-order Butcher's Runge-Kutta method (RK5) that
makes 6 function evaluations over one integration step. We used moderate
integration times such as $t_{\max}\approx6000$ with $SJ\Rightarrow1$
and $\hbar\Rightarrow1$, as usual, and at the end computed the fluctuation
spectrum (FS) of the quantity $F$ as follows:
\begin{equation}
\mathrm{FS}(\omega)=\frac{1}{t_{\max}}\left|\intop_{0}^{^{t_{\max}}}F(t)e^{i\omega t}dt\right|^{2}.\label{FS}
\end{equation}
As the quantity $F$ we used either the $X$ displacement of the skyrmion
(or the average displacement of skyrmions in the SkL) in the TM subsystem,
$X_{S}$, from its equilibrium position or the $x$ component of the
magnetic moment of the system. The modes' frequencies were searched
for as the resonances in $\mathrm{FS}(\omega)$. We use the system
size $116\times132$ in lattice units (with periodic boundary conditions)
that neatly comprises an SkL of 12 skyrmions, as shown in Fig. \ref{Fig_SkL}.

\begin{figure}
\centering{}\includegraphics[width=8cm]{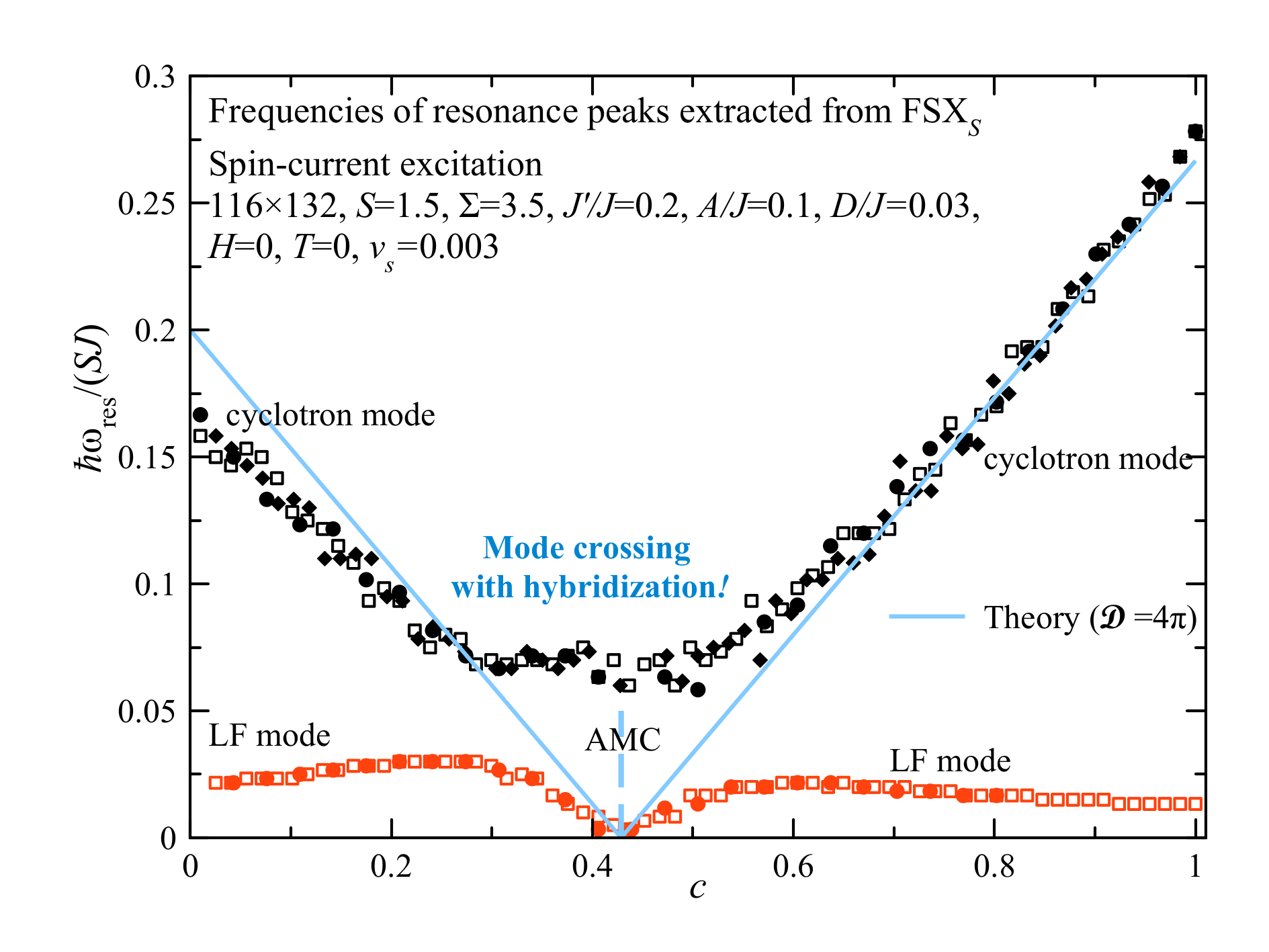} \caption{The dependence of the hybridized SCR and LF ferrimagnetic modes on
the concentration of Gd in a model of CoGd ferrimagnet. (Single skyrmion
in a $116\times132$ system, spin-current excitation, and the TM skyrmion
position, $X_{S}$, as the observed quantity. The results for the
SkL are similar). }\label{Fig_omega_res}
\end{figure}

The dependence of the hybridized SCR and LF ferrimagnetic modes on
the concentration of Gd $c$ for a single skyrmion (with the size
$\lambda_{\mathrm{eff}}/a\approx18)$ in the system in the case of
spin-current excitation is shown in Fig. \ref{Fig_omega_res}. As
the result of hybridization, there are upper and lower modes with
a gap between them. Far from the angular-momentum compensation (AMC)
point, $c^{*}=S/\Sigma=0.429$, the upper mode is the SCR mode, and
the lower mode is the LF ferrimagnetic mode. In the region around
the AMC point, the lower mode becomes mainly the SCR mode with the
frequency going to zero at $c=c^{*}$. For small $c$, the peaks in
$\mathrm{FS}(\omega)$ corresponding to both upper and lower modes
become very small and finally disappear in the limit $c\rightarrow0$.
We have separately checked the proportionality of $\Omega$ to $J'$
at a constant $c$ far from compensation. Excellent quantitative agreement
with the analytical formula (\ref{Omega_cyclotron}) is apparent,
as is the expected qualitative dependence of both modes on $c$ due
to their hybridization. 

The frequencies of the modes can also be obtained by computing the
energy absorbed at different frequencies for a given $c$. There are
two absorption peaks corresponding to the upper and lower modes. With
an eye on experiments, one should worry about the power absorbed by
the cyclotron mode. A single skyrmion can hardly provide a detectable
signal, so one should consider the power absorption by a dense skyrmion
lattice created in a ferrimagnet. The modes' frequencies for the SkL
obtained via $\mathrm{FS}(\omega)$, are rather close to the results
of Fig. \ref{Fig_omega_res}.

A similar computation with the microwave excitation reproduces the
lower mode, while the upper mode cannot be reliably resolved. In these
experiments, the observed quantity $F$ in Eq. (\ref{FS}) can be
both the skyrmion displacement $X_{S}$ and the transverse magnetization
component. The latter is easier to detect experimentally. These observations
point toward the possibility to obtain the evidence of the skyrmion
cyclotron motion in the microwave experiment as well. 

\begin{figure}
\centering{}\includegraphics[width=8cm]{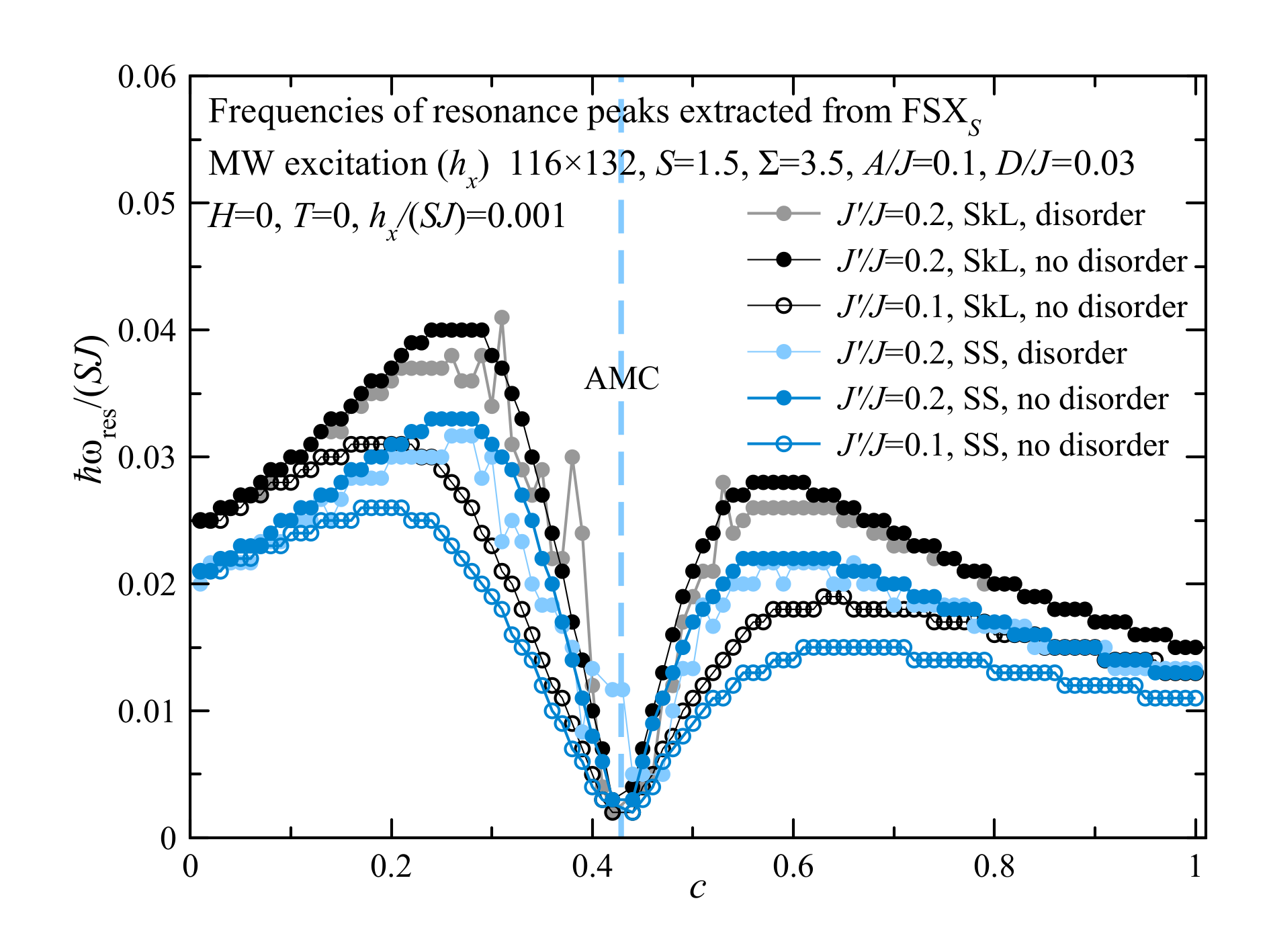} \caption{The dependence of the lower mode on the concentration of Gd for a
single skyrmion (SS) and the skyrmion lattice in a model of CoGd ferrimagnet.}\label{Fig_omega_res_lower}
\end{figure}

Figure\ \ref{Fig_omega_res_lower} shows the RE concentration dependence
of lower mode's frequency in the case of MW excitation for a single
skyrmion (SS) and the SkL. In both cases, there is a spectacular depression
near the AMC point. The curves labeled \textquotedbl no disorder\textquotedbl{}
were obtained for the idealized model in which the length of each
RE spin is $c\Sigma$. These curves do not have the scatter caused
by the disorder. For smaller intersublattice coupling, $J'/J=0.1$,
the depression is expectedly broader. 

In conclusion, we have demonstrated that a universal formula for the
skyrmion mass in a ferrimagnet corresponds to a universal formula
for the frequency of the gyroscopic motion of the skyrmion. Such a
motion of skyrmions in ferrimagnets must show as a skyrmion cyclotron
resonance, SCR, in the absorption of the power of the ac spin current,
and in the hybridization with ferrimagnetic modes when skyrmions are
present, which must be possible to detect in microwave experiments
as well. Similar to the electron cyclotron resonance in metals, observation
of the SCR would allow experimentalists to unambiguously measure the
skyrmion mass which has been subject of significant controversy. 

Our analytical formulas equally apply to synthetic ferrimagnets in
which antiferromagnetically coupled sublattices belong to different
adjacent ferromagnetic layers \citep{Soum-Nat2017,Xia-PRB2021,Mallik-Nat2024}.
The numerical results are expected to be similar for such systems.
They provide another possibility for experimental detection of the
skyrmion cyclotron resonance and for obtaining the skyrmion mass.

Finally, we would like to notice the possibility of quantization of
the skyrmion cyclotron motion, similar to Landau quantization of an
electron's orbital motion in a magnetic field.

This work has been supported by Grants No. FA9550-24-1-0090 and FA9550-24-1-0290,
funded by the Air Force Office of Scientific Research.\\

\end{document}